\documentclass[twocolumn]{aastex61}

\usepackage{amsmath,amstext}
\usepackage[T1]{fontenc}
\usepackage[figure,figure*]{hypcap}

\shorttitle{}
\shortauthors{Lovekin \& Guzik}

\begin{document}

\title{Convection and Overshoot in models of $\gamma$ Doradus and $\delta$ Scuti stars}

\author{C.C. Lovekin}
\affiliation{Physics Department, Mount Allison University, Sackville, NB, E4L 1C6, Canada}
\affiliation{Kavli Institute for Theoretical Physics, University of California,
Santa Barbara, Santa Barbara, California 93106, USA}
\email{clovekin@mta.ca}
\author{J.A. Guzik}
\affiliation{XTD-NTA, MS T-082 Los Alamos National Laboratory, Los Alamos, NM, 87545, USA}
\affiliation{Kavli Institute for Theoretical Physics, University of California,
Santa Barbara, Santa Barbara, California 93106, USA}

\begin{abstract}
We investigate the pulsation properties of stellar models representative of $\delta$ Scuti and $\gamma$ Doradus variables.  
We have calculated a grid of stellar models from 1.2 to 2.2 M$_{\odot}$, including the effects of both rotation and convective overshoot using MESA, and we investigate the pulsation properties of these models using GYRE.  
We discuss observable patterns in the frequency spacing for $p$ modes and the period spacings for $g$ modes.  Using the observable patterns in $g$ mode period spacings, it may be possible to observationally constrain the convective overshoot and rotation of a model.  
We also calculate the pulsation constant (Q) for all models in our grid, and investigate the variation with convective overshoot and rotation.  The variation in Q values of radial modes can be used to place constraints on the convective overshoot and rotation of stars in this region.  As a test case, we apply this method to a sample of 22 high amplitude $\delta$ Scuti stars (HADS), and provide estimates for the convective overshoot of the sample.  
\end{abstract}


\section{Introduction}

$\delta$ Scuti stars are main-sequence stars with masses 1.4 - 2.5 times the mass of the Sun, located where the classical instability strip intersects the main sequence in the HR diagram.  The pulsations are driven by the $\kappa$ mechanism in the HeII ionization zone, and are either low order pressure modes ($p$-modes) \citep{xiong2016}, or mixed modes with both $p$ and $g$ mode properties \citep{guzik2000,Dupret2004,dupret2005}.  The observed pulsation periods are usually short, ranging from 30 minutes to 5 hours.  Observations using the \emph{Kepler} spacecraft \citep{basri} have shown that $\delta$ Scuti stars have very rich pulsation spectra with many identifiable frequencies. 

$\delta$ Scuti stars that pulsate with a peak-to-peak amplitude greater than about 0.3 mag are known as High-Amplitude $\delta$ Scuti stars (HADS).  This sub-group is generally thought to be slowly rotating ($v \leq 30 km/s$), and are thought to pulsate mostly in radial modes.  Given their location in the HR diagram, these stars have been viewed as intermediate objects, located between the normal $\delta$ Scuti stars and Cepheids.  However, \cite{balona2016} find that there is no difference in distribution between HADS and low amplitude $\delta$ Scuti stars, suggesting that these two groups of stars are drawn from the same population. However, HADS account for a small fraction of the total population \citep{lee2008}, around 0.24 \%.  

One of the theoretical challenges presented by $\delta$ Scuti stars is determining which stars should be pulsating.  Although all the stars in the $\delta$ Scuti instability strip are predicted to show pulsations, observations show that less than half of the stars in this region actually pulsate \citep{balona2015,balona2011}.  This difference between theory and observation is not presently understood.  It has been suggested that nonlinear mode coupling may stabilize the pulsations \cite{dziembowski88}, or the opacity driving mechanism could be saturated \citep{nowakowski05}.  Unfortunately, testing these predictions requires nonlinear models of nonadiabatic nonradial pulsations, which do not currently exist.

$\gamma$ Doradus stars are slightly less massive than $\delta$ Scuti stars, with an instability region that overlaps the cool part of the $\delta$ Scuti instability strip. The $\gamma$ Dor stars pulsate in $g$ modes, which are thought to be triggered by convective blocking \citep{guzik2000,Dupret2004,dupret2005}.  The periods are longer than in $\delta$ Scuti stars, ranging from 0.3 to 3 days.  The low amplitudes and pulsation periods of order 1 day make $\gamma$ Doradus stars difficult to observe from the ground.  Space-based observations, such as with COROT \citep{hareter2010} and \emph{Kepler} \citep[e.g.,][]{grigahcene2010,bradley2015} have found that a number of $\delta$ Scuti stars are observed to pulsate with $\gamma$ Dor-type frequencies as well, and these are usually described as hybrid stars.  Hybrid stars are particularly interesting asteroseismic targets, as the $g$ modes sample deeper regions of the star than the $p$ modes.  Hybrid stars, $\gamma$ Dor and $\delta$ Scuti stars collectively span the mass range 1.4-2.5 M$_{\odot}$, where the interior structure transitions from a convective envelope and radiative core in the lower mass stars to higher mass stars with convective cores and radiative envelopes.  Studies of stars in this mass range with \emph{Kepler} have found more hybrid star candidates than expected by current stellar pulsation theory \citep{uytterhoeven2011}.  The leading explanation for this discrepancy is that the interaction between convection and pulsation in both the deeper layers ($\gamma$ Dor, $g$ modes) and upper layers ($\delta$ Scuti, $p$ modes) is not fully understood.  

Recent observations by \citet{balona2014} and \citet{balona2015} have suggested that the low frequencies typical of $\gamma$ Dor variables are found in all $\delta$ Scuti stars.  They argue that, from the ground, only the higher-amplitude $\delta$ Scuti modes are visible, but in fact, all pulsating stars in this region are ``hybrid''.  If true, this would only deepen the discrepancy between theoretical predictions and observed pulsation.  Internal convection, both in the core and sub-surface, is likely to play an important role in determining the pulsation properties of these stars.  For models to correctly predict pulsations, the interaction between convection and pulsation must be included.

Rotation is also an important factor in $\delta$ Scuti stars, as these stars can have rotation rates as high as 200-250 km/s \citep{breger2007}.  Rotation can affect the small-scale features of convection, in particular the dynamics of the shear layers at the boundaries of convective zones.  The details of these small-scale features are likely to be a key part in correctly predicting pulsation frequencies in these rapidly rotating stars.  On larger scales, rotation can increase the size of the convective core, as parametrized in our stellar models by the convective core overshoot.  

Many types of variable stars are found to obey a period-mean density relation \cite[see, for example][]{cox1980}.  This relationship is usually written as
\begin{equation}
\Pi\left(\frac{\overline{\rho}}{\overline{\rho_{\odot}}}\right)^{1/2} = Q.
\end{equation}
Q, known as the pulsation constant, is approximately constant for stars that are similar in size and structure.  The pulsation constant is expected to apply to $\delta$ Scuti pulsations, both radial and non-radial.  Relationships between the pulsation constant and observable properties of the star have been determined by \cite{breger1990}, and these can be used to identify an observed mode in these stars.  However, this relationship is calibrated to the Sun, and neglects the effects of rotation and convective overshoot on the pulsation.

The behaviour of Q has previously been studied in $\delta$ Scuti stars by \citet{Fitch81}.  In this work, \citeauthor{Fitch81} studies models of 1.5, 2.0 and 2.5 M$_{\odot}$, and pulsation periods were calculated for $\ell \leq 3$.    Based on these models, \citeauthor{Fitch81} calculated Q for the first 7 overtones, and then determined the coefficients for interpolation formulae.  These formulae can be used to predict the expected value of Q for a given mode based on the observed period and effective temperature of a star.

In this work, we investigate the pulsation characteristics of stellar models in the range 1.2 - 2.2 $M_{\odot}$.  Our model grid, calculated using MESA \citep{mesa} and GYRE \citep{GYRE}, is described in Section \ref{sec:models}.  In Section \ref{sec:Q}, we discuss the pulsation constant (Q) of the models, and the potential for using Q to constrain the core convective overshoot and rotation of populations of stars.  We present the patterns in the frequency and period spacings in Section \ref{sec:patterns}.  Finally, we summarize our conclusions in Section \ref{sec:conclusions}.  

\section{Models}
\label{sec:models}

We calculate a grid of stellar evolution models using MESA \citep{mesa} for masses from 1.2 to 2.2 M$_{\odot}$.  These models were calculated using metallicity of $Z = 0.02, Y = 0.28$, from \citet{GS98}.  MESA uses the OPAL equation of state \citep{OPAL}, supplemented with the SCVH tables \citep{SCVH} at low temperatures and densities.  Convection is treated using the mixing length theory (MLT) formalism \citep{MLT} with $\alpha$ = 2.0.

Our models include varying rotation and convective overshoot.  Overshoot is included using the exponential model of \cite{Herwig2000}, where the diffusion coefficient is given by
\begin{equation}
D_{ov} = D_0exp\left(\frac{-2r}{f_{ov}H_P}\right)
\end{equation}
where $D_0$ is the diffusion coefficient at the convective boundary, $r$ is the radial distance from the core boundary, and $H_P$ is the pressure scale height at the core boundary.  The amount of overshooting is defined in terms of a fraction of the pressure scale height, $H_P$.  The fraction is parameterized using $f_{ov}$, the overshooting parameter.  The overshoot parameter $f_{ov}$ is allowed to vary between 0 and 0.1 in our models.  Overshooting is permitted both above convective cores and below convective envelopes.  

Rotation is included by imposing a surface rotation rate and forcing the zero-age main sequence (ZAMS) model to be uniformly rotating.  The surface rotation rate ($\Omega$) is defined as a fraction of critical rotation ($\Omega_c$), or the point at which the gravitational force is balanced by the centripetal acceleration at the equator.   We parametrize the rotation rate as $\Omega/\Omega_c$, and allow it to vary between 0 and 0.6 in our models.  MESA uses the shellular treatment of rotation \citep{meynet1997}, which does not take into account the centrifugal deformation of the star.  Although this 1D treatment breaks down at very high rotation rates, it is sufficient for the rotation rates discussed here.  Transport of angular momentum is implemented via a diffusion approximation \citep{MESA2}.  

Pulsation frequencies for each model were calculated using GYRE \citep{GYRE} with frequencies ranging from 2 to 400 $\mu$Hz (0.17-34.56 c/d) with $\ell$ = 0, 1, and 2 for all allowed values of $m$.  This covers radial orders of $g$ modes from 1 to approximately 100 depending on the rotation rate, and the first few radial orders of the $p$ modes.  We calculated non-adiabatic frequencies, giving us estimates for the growth rates of all calculated modes.  GYRE includes rotation using the Traditional Approximation.  In \cite{prat2017}, the authors compare the Traditional approximation to a more accurate ray-tracing calculations, and find that the resulting period spacings are accurate to high rotation rates, at least for low-frequency g-modes.  For p-modes, the traditional perturbation methods have been shown to be reliable up to about 0.4 $\Omega/\Omega_c$ \citep{lignieres2006, lovekin2008}. Above this limit, frequencies can differ by more than 10\%, depending on the radial order and $\ell$ of the mode.

\section{Pulsation Constants}
\label{sec:Q}
\begin{figure}
\includegraphics[width=0.5\textwidth]{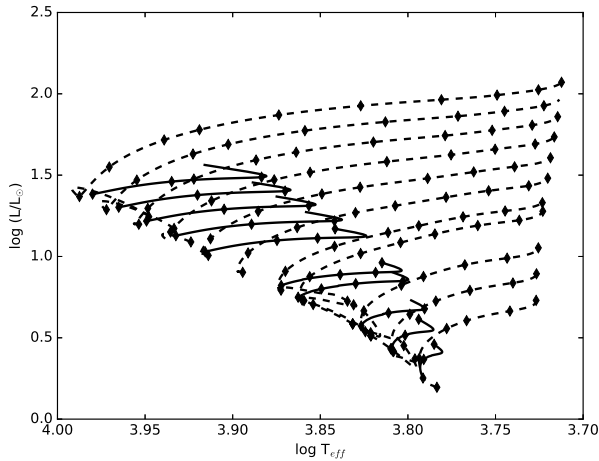}
\caption{\label{fig:HR}Sample evolution tracks for stellar models ranging from 1.2 to 2.1 M$_{\odot}$.  Models evolved with no overshoot ($f_{ov} = 0$) are shown with solid lines, while the models with $f_{ov} = 0.1$ are shown with dashed lines.  The individual models used for the calculation of pulsation constants are indicated with diamonds.  These two sets of tracks cover the extremes of our parameter space, with other overshoot parameters and rotation rates falling within this range in logT$_{eff}$ and log$L/L_{\odot}$.}
\end{figure}

\begin{figure}[!ht]
\includegraphics[width=0.5\textwidth]{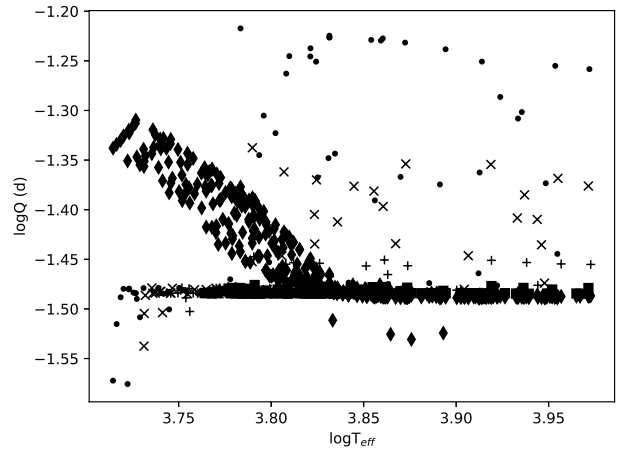}
\caption{\label{fig:Q_l=0}The calculated Q values based on the MESA/GYRE models for the radial $n = 1$ mode for all non-rotating models in our grid (diamonds).  The \cite{Fitch81} scaling law predictions, shown with squares, +, $\times$, and $\bullet$,  agree well for $f_{ov} \leq 0.04$ (squares) for temperatures greater than log$T_{eff} \approx 3.83$.  A small amount of scatter is seen at overshoot $f_{ov} = 0.06$ (+), larger scatter at $f_{ov}= 0.08$ ($\times$), and the largest amount of scatter corresponds to the largest overshoot ($f_{ov} = 0.1, \bullet$).  Below logT$_{eff} \approx 3.83$, we see an increasing trend in Q that is not seen in \citet{Fitch81}.  }
\end{figure}

We have calculated Q values for all p-modes with $n \leq 7$ for the models in our grid, and compared our calculated Qs to the Q values predicted by \cite{Fitch81}'s interpolation formulae.  For each evolution sequence, we selected every 20th model along the evolution sequence.  Sample tracks are shown in Figure \ref{fig:HR} for models with $f_{ov}$ = 0 (solid) and $f_{ov}$ = 0.1 (dashed), spanning the range of our parameter space.  The models used to calculate the pulsation constants are indicated by the diamonds. For the radial n = 1 modes shown in Figure \ref{fig:Q_l=0}, the \cite{Fitch81} predictions agree well for models with log$T_{eff} \gtrapprox 3.83$, and none were cooler than logT$_{eff} = 3.80$. Below this temperature, the parametrization predicts much lower values, as our models show a significant change in the slope of log$Q$ as a function of log$T_{eff}$.  This change in slope was not accounted for by \cite{Fitch81}.  In the original parametrization, the predicted Q depends only on the period, and is independent of temperature.  Indeed, only two models in their grid had a temperature below log$T_{eff} \approx 3.83$, where we find the change in slope for the radial $n = 1$ mode.  In the temperature range spanned by \citet{Fitch81}, the Q value calculated from our models is also independent of the effective temperature.  We also find that given the parameters of our models, the parametrization of \citeauthor{Fitch81} produces a much larger scatter in Q than we observe based on our calculations.  This scatter exists only in models with $f_{ov} \geq 0.06$, and increases dramatically as a function of overshoot.  Increasing overshoot tends to increase the pulsation period in our models.  The scaling relations presented by \citet{Fitch81} for the $\ell = 0$ modes depend only on the period, which means that these relations will tend to predict a higher Q.  Since overshoot also decreases the mean density, this increase in Q is not seen in our models. The higher the overshoot, the more pronounced this effect becomes, resulting in higher predicted Q parameters at higher values of convective overshoot.

\begin{figure}
\includegraphics[width=0.5\textwidth]{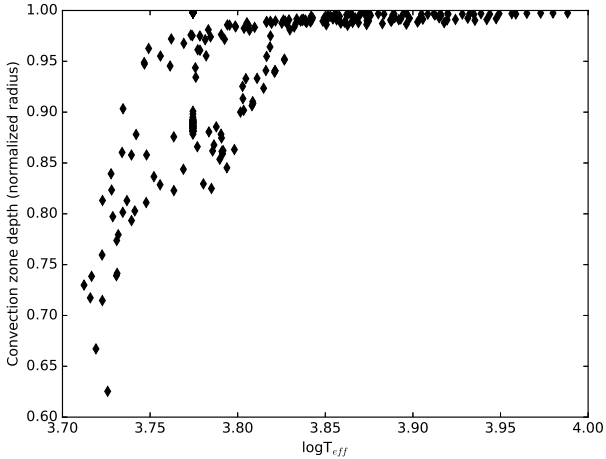}
\caption{\label{fig:convectionzone}The normalized depth of the convection zone as a function of effective temperature for all models in our grid.  The convective envelope begins to grow as models become cooler than logT$_{eff}$ = 3.83.  This boundary corresponds to the change in slope seen in the logQ-logT$_{eff}$ plots (see, for example Figure \ref{fig:Q_l=0}).  }
\end{figure}

The slope in Q below logT$_{eff} \approx 3.83$ seen in \ref{fig:Q_l=0} is related to the depth of the convective envelope.  In Figure \ref{fig:convectionzone} we show the fractional depth of the convective envelope as a function of effective temperature.  In models hotter than logT$_{eff}$ = 3.83, the convective envelope remains very shallow.  In models that are either initially cooler than logT$_{eff}$ = 3.83 or evolve to lower temperatures, the convective envelope beings to grow, extending deeper into the star.  The change in temperature and sound speed gradient changes the frequencies, resulting in the observed slope in logQ.

\begin{figure}[!ht]
\includegraphics[width=0.5\textwidth]{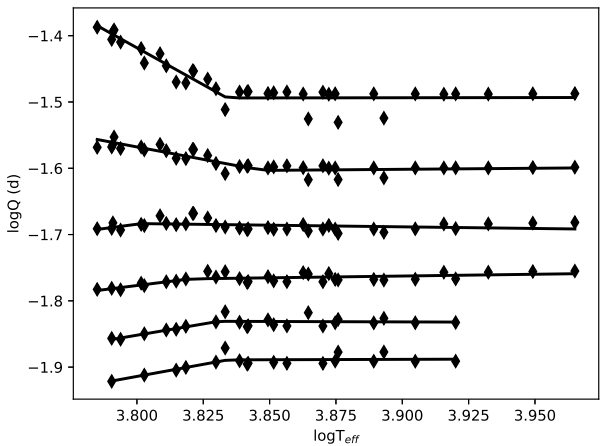}
\caption{\label{fig:Q_alln}Calculated Q values for $\ell = 0$, $n$ = 1 (top) to 7 (bottom) for all non-rotating, zero overshoot models in our grid. $n$ decreases downwards in this plot.  The solid lines show the piecewise linear fit to the data.   }
\end{figure}

As the radial overtone of the modes increases, we find that the change in slope becomes less sharp, and actually changes sign at $n \approx 4$, as shown in Figure \ref{fig:Q_alln} for the $\ell = 0$ modes.  In all cases, the change in slope is found near log$T_{eff} = 3.83$.  On the cool side of this break are the models with $M \leq 1.4 M_{\odot}$ and the later stages of main sequence evolution for the more massive models.

We fit each n value separately with a piecewise linear function.  
\begin{eqnarray}
logQ &= &m_{low}logT_{eff} + b_1 \quad logT_{eff} < T_{break} \nonumber \\
log Q& =& m_{high}logT_{eff} + b_2 \quad logT_{eff} \geq T_{break}
\end{eqnarray}
with the restriction that $b_1$ = $(m_{low}-m_{high})T_{break} + b_2$.  
The four free parameters are the break point, $T_b$, the slope for $x < x_b$ ($m_{low}$), the slope for $x \geq x_b$ ($m_{high}$), and the intercept of the fit ($b_2$).  These fits are also shown on Figure \ref{fig:Q_alln}.  The parameter $m_{high}$ was not found to vary significantly with any of the parameters, and is very close to 0 in all cases.  For this reason, we focus on the other three parameters below. We used the best fitting values of $m_{low}$, $T_{break}$, and $b2$ to investigate the effects of rotation and overshoot on log$Q$.  

We found that the parameters $b_2$ and $T_b$ were basically constant as the overshoot was increased.  Increasing overshoot does increase $m_{low}$, with the exception of the $f_{ov}$ = 0 models.  However, this particular set of models shows a higher amount of scatter, particularly near the break point.  This scatter appears to artificially decrease the slope, which should in fact be steeper if these points were excluded.  If we consider only the parameter $m_{low}$, over the range $f_{ov}  0.02 - 0.1$, the variation in $m_{low}$ as a function of overshoot can be well fit by a straight line, with 
\begin{equation}
m_Q = 14 f_{ov} - 2.8
\end{equation}
for the radial $n = 1$ mode, where $m_Q$ is the predicted value of $m_{low}$.  We found that for a given $n$, there was not a significant variation in the slopes ($m_{low}$) with rotation.  At $\ell= 0$, the exception is that the Q values for models with $\Omega/\Omega_c = 0.5$ are significantly lower than the other rotation rates.  To fit $m_{low}$, we averaged the Q values at each overshoot over all rotation rates and fit the averages, excluding the Qs for the models with $\Omega/\Omega_c = 0.5$.  The variation in the resulting $m_{low}$ as a function of overshoot can be fit by 
\begin{equation}
m_Q = (12\pm 2) f_{ov} - (2.8 \pm 0.1)
\label{slopefit}
\end{equation}
for the radial $n = 1$ mode.  
We fit the variation in this slope for the first 7 radial orders of the $\ell= 0 $ modes in our models. The mean and standard deviation for $m_{ov}$, the slope of the $m_Q - f_{ov}$ relationship, and for $b_{ov}$, the intercept of this relationship are summarized in Table \ref{tab:Qfits}. 

\begin{deluxetable}{ccccc} 
\tabletypesize{\footnotesize}  
\tablecaption{\label{tab:Qfits} Trends in the slope of log$Q$ vs log$T_{eff}$ ($m_{low}$) as a function of overshoot} 
\tablehead{ 
\colhead{$n$} & \colhead{$m_{ov}$} & \colhead{$\sigma(m_{ov})$} & \colhead{$b_{ov}$} & \colhead{$\sigma(b_{ov})$} \\}
\startdata 
1 & 12 & 2 & -2.8 & 0.1 \\
2 & 3.9 & 0.6 & -0.59 & 0.04 \\
3 & -0.43 & 0.5 & 0.17 & 0.04 \\
4 & -1.7 & 0.6 & 0.48 & 0.04 \\
5 & -2.2 & 0.7 & 0.67 & 0.02 \\
6 & -3.1 & 0.8 & 0.84 & 0.03 \\
7 & -5.5 & 0.5 & 1.02 & 0.02 \\
\enddata  
\end{deluxetable}

Similarly, we note that the intercept of the fit ($b_2$) appears to be sensitive to the amount of rotation in the models. The amount of convective overshoot does not strongly affect this part of the plot, so we average over all overshoot values when fitting.  As for the variation in $m_{low}$, the variation in $b_2$ shows a linear variation with rotation rate, although the scatter at high rotation rates ($\Omega/\Omega_c \geq 0.5$) is large.  We fit $b_2$ as a function of rotation rate with a straight line, and present the coefficients of the fit (the slope, $m_{\Omega}$ and the intercept, $b_{\Omega}$) at each $n$ in Table \ref{tab:levelfits}.  We also find some indications that there is a relationship between the break point in our fit to the Q data and the rotation rate of the models.  However, the scatter in this relationship is high, and the variation in the break point is small. For this reason, we do not present fits to the trends here.

\begin{deluxetable}{ccccc} 
\tabletypesize{\footnotesize}  
\tablecaption{\label{tab:levelfits} Trends in the intercept ($b_2$) of our piecewise linear fit as a function of rotation rate.} 
\tablehead{ 
\colhead{$n$} & \colhead{$m_{\Omega}$} & \colhead{$\sigma(m_{\Omega})$} & \colhead{$b_{\Omega}$} & \colhead{$\sigma(b_{\Omega})$} \\}
\startdata 
1 & -0.8 & 0.2 & -1.14 & 0.07 \\
2 & 0.22 & 0.06 & -1.60 & 0.02 \\
3 & 0.2 & 0.2 & -1.66 & 0.03 \\
4 & 0.8 & 0.1 & -2.15 & 0.03 \\
5 & 0.47 & 0.03 & -2.03 & 0.02 \\
6 & 0.57 & 0.09 & -2.09 & 0.03 \\
7 & 0.6 & 0.1 & -2.18 & 0.03 \\
\enddata
\end{deluxetable}

In principle, this result could be used to estimate the average rotation and overshoot for populations of stars.  We have tested this method on a small sample of high amplitude $\delta$ Scuti stars (HADS) taken from the International Variable Star Index maintained by the American Association of Variable Star Observers (AAVSO).  We collected a sample of 22 stars known to pulsate in radial modes.  The basic parameters of these stars are presented in Table \ref{tab:stars}.  We estimated effective temperatures for these stars using the B and V magnitudes available in the SIMBAD database and the color-effective temperature relationship determined by \cite{Sekiguchi}.  We used the period of the highest-amplitude mode in the AAVSO catalog as the radial $n = 1$ mode, and calculated Q assuming that all stars had the same mean density. The resulting effective temperatures and Q values are unlikely to have very high accuracy, but should be sufficient for a proof-of-concept test of our method.  

\begin{figure}
\includegraphics[width=0.5\textwidth]{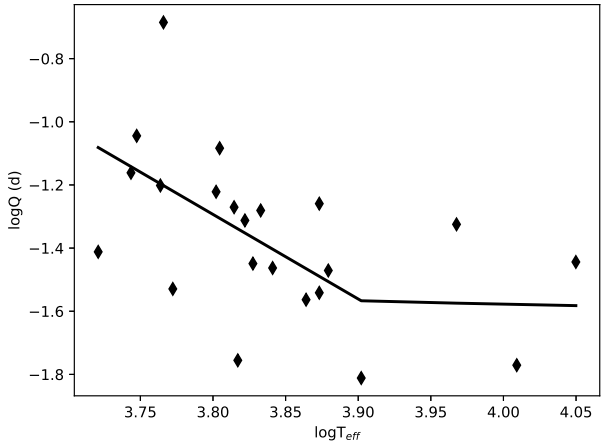}
\caption{\label{fig:testcase}The logQ-log$T_{eff}$ diagram for the 22 HADS in our sample.  The observed stars are marked by diamonds, and are fit with a broken power law (solid line).  The same trend is seen in the observed stars as in the model data.  The slope of the line for stars with log$T_{eff} < 3.85$ is -2.7 $\pm$ 1.2.}
\end{figure}

We plotted logQ versus log$T_{eff}$ for each of the 22 stars in our sample, and fit them with the same broken power law fit we used to fit our models.  The result is shown in Figure \ref{fig:testcase}.  The broken power law used to fit the model data also appears to fit the data in this case, although the small number of observations leads to large uncertainties.  We find the observed slope of the cool stars is -2.7 $\pm$ 1.2, and we can use this value in Equation \ref{slopefit} to estimate the convective overshoot of the stars in our sample.  For these stars, we estimate that the convective overshoot $f_{ov} = 0.008_{-0.008}^{+0.4}$.  The errors on this estimate are of course extremely large, but it should be possible to reduce the errors by using a larger sample of stars with more accurate measurements of effective temperature and mean density.

We are also able to determine the average logQ for the stars in this sample hotter than log$T_{eff} = 3.85$ is -1.2 $\pm$ 8.3.  However, the sample presented here has very few ($\approx 4$) stars hotter than log$T_{eff} = 3.85$, so the estimate has extremely high uncertainty.  Again, a larger sample of better constrained stars should make more accurate estimates of the rotation rate possible.  This is a promising area of future work.  In particular, for this technique to be practical for application to observed stars, we will need to investigate the dependence of the scaling relations on metallicity and helium abundance.  It would also be interesting to see if these same relations hold for other radial pulsators, such as classical Cepheids or RR Lyrae variables.

\begin{deluxetable}{lccc}
\tablecaption{\label{tab:stars}The 22 stars used to test the method for determining rotation and overshoot described above.}
\tablehead{\colhead{Identifier} & \colhead{Period (d) }  & \colhead{(B-V)} & \colhead{log$T_{eff}$} }
\startdata
AI Vel & 0.11  & 0.28 & 3.84 \\
BPS BS 16553-0026 & 0.13  & 0.82 & 3.72 \\
NSV 14800 & 0.16  & 0.35 & 3.82 \\
USNO-B1.0 0961-0254829 & 0.05  & -0.20 & 4.01 \\
V0703 Sco & 0.12  & 0.33 & 3.83 \\
BP Peg & 0.11  & 0.15 & 3.88 \\
2MASS J18294745+3745005 & 0.12  & -0.29 & 4.05 \\
VX Hya & 0.22  & 0.70 & 3.74 \\
V0879 Her & 0.06  & 0.37 & 3.82\\ 
NSVS 7293918 & 0.09  & 0.20 & 3.86\\ 
V0403 Gem & 0.15  & -0.10 & 3.97 \\
GSC 03949-00811 & 0.17  & 0.31 & 3.83 \\
GSC 03949-00386 & 0.10  & 0.56 & 3.77 \\
ASAS J194803+4146.9 & 0.20  & 0.60 & 3.76 \\
V1719 Cyg & 0.27  & 0.42 & 3.80 \\
GSC 04257-00471 & 0.17  & 0.38 & 3.81 \\
V0823 Cas & 0.67  & 0.59 & 3.77 \\
DO CMi & 0.19  & 0.43 & 3.80 \\
VZ Cnc & 0.18  & 0.17 & 3.87 \\
2MASS J06451725+4122158 & 0.05  & 0.08 & 3.90 \\
RV Ari & 0.09  & 0.17 & 3.87 \\
V0829 Aql & 0.29  & 0.68 & 3.75 \\
\enddata
\end{deluxetable}

\begin{figure}
\includegraphics[width=0.5\textwidth]{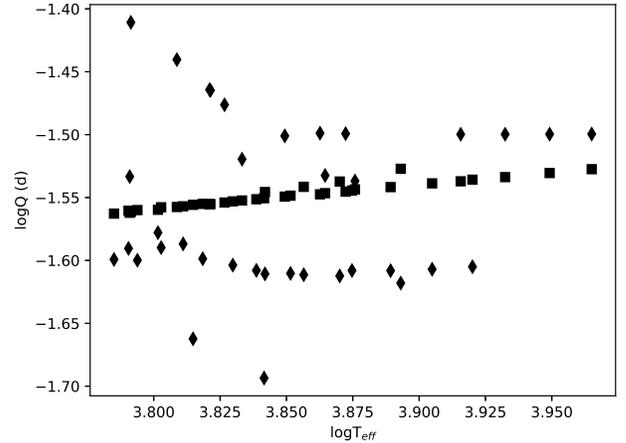}
\caption{\label{fig:nonradQ}The logQ-log$T_{eff}$ diagram for $\ell = 1, m = 0$ modes in non-rotating models with no overshoot.  As in Figure \ref{fig:Q_l=0}, Q values for our models are shown with diamonds, while Q values calculated from the fitting formulae of \cite{Fitch81} are marked with squares.}
\end{figure}

We have also calculated the Q values for modes with $\ell = 1$ and 2 and compared them to the predictions from \cite{Fitch81}.  In the non-radial case, \citeauthor{Fitch81}'s predictions do not work as well, as illustrated in Figure \ref{fig:nonradQ} for $\ell = 1$ modes. Our model Q values show a bimodal distribution.  The lower curve (shorter Q) consists of main sequence models, while models in the blue hook phase of evolution are on the upper sequence.  \citeauthor{Fitch81}'s predictions are based on subgiant models, and lie between the two sequences.  For the $\ell = 2$ models, we find that the \cite{Fitch81} predictions over predict the expected logQ by between 0.1 and 0.2 dex.  For both the $\ell = 1$ and $\ell = 2$ models, there are no significant trends with either rotation or overshoot, so these modes cannot be easily used to constrain these parameters.  

\section{Regular Patterns in Observed Pulsation}
\label{sec:patterns}
\subsection{Frequency Spacings}

The $p$-mode pulsations in $\delta$ Scuti stars are of low radial order, and are not expected to obey the asymptotic frequency spacing relations of \citet{tassoul80}.  However, regular frequency spacings have been observed in at least some $\delta$ Scuti stars \citep[eg.][]{handler97,breger99}, and various techniques have been proposed for identifying the patterns \citep{garciahernandez2013,paparo2016}.  Identification of frequency patterns can be quite challenging though, as \citet{paparo2016} find that in some cases, as few as 20\% of the frequencies can be located on the echelle ridges.  

Since the mode identification for our models is known, it is a straightforward matter to construct echelle diagrams and to determine the large frequency spacing ($\delta \nu$).  We have found that increased rotation and convective overshoot both increase the large frequency spacing, although $\delta \nu$ increases more rapidly with rotation than with convective core overshoot.  Both rotation and convective core overshoot introduce extra mixing, which will tend to smooth out composition gradients surrounding the convective core, making the star appear younger.  This is consistent with the results found by \cite{JCD2005}, which shows that the large separation increases as the star evolves.  Unfortunately, we have no method for disentangling the effects of rotation and overshoot, so although a larger than expected $\delta \nu$ indicates the effects of convective overshoot and/or rotation are present, we cannot use the echelle diagrams and the large separation to quantify the processes.  

\subsection{Period Spacings}

\begin{figure}
\includegraphics[width=0.5\textwidth]{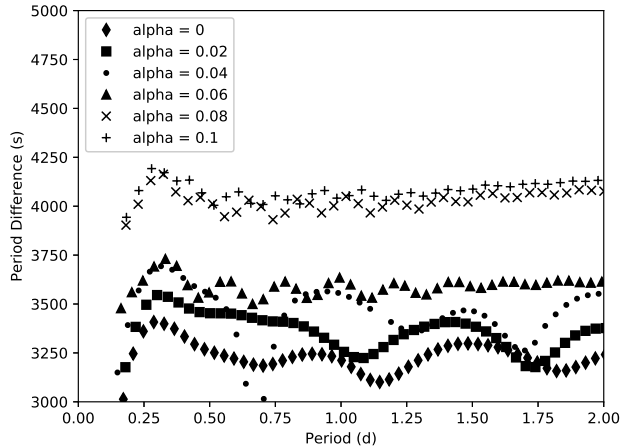}
\caption{\label{fig:alphaspacing}The period spacing as a function of period for a 1.6 M$_{\odot}$ model with no rotation.  Increased mixing from convective overshooting smooths out the chemical discontinuity surrounding the core, smoothing the period spacings.  Increased overshoot also increases the absolute value of the period spacing.   }
\end{figure}

The $\gamma$ Doradus variables are expected to display regular patterns in the period spacing \citep{vanreeth15,vanreeth16,ouazzani2016}.  As found by these previous studies, we find that in non-rotating models, the period spacings are nearly flat as a function of period, as shown in Figure \ref{fig:alphaspacing}.  Adding rotation lifts the degeneracy of the $m$ values, so for $m = 1$ modes the period spacings have a negative slope, while for $m = -1$ the period spacings have a positive slope.  As the age of the model increases, the chemical discontinuity at the core causes regular dips in the period spacings.  

Convective overshoot has two effects on the predicted period spacings.  First, as noted by \citet{vanreeth15}, the increased mixing caused by convective overshooting tends to wash out the chemical discontinuity, so the dips in the period spacing pattern become smaller, as shown in Figure \ref{fig:alphaspacing}.  We also find that increased convective overshoot increases the absolute value of the period spacing.  Increased convective overshooting effectively increases the size of the hydrogen burning core, so the star behaves as though it were more massive.  This trend for increasing period spacing is identical to that seen as mass increases.  At the highest overshoots in our grid $f_{ov} = 0.1$, the increase in period spacing is the same as that produced by increasing the mass by approximately 0.8 M$_{\odot}$.  

\begin{figure}
\includegraphics[width=0.5\textwidth]{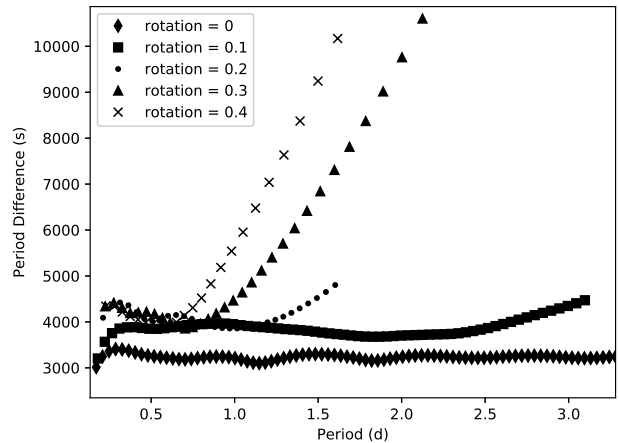}
\caption{\label{fig:rotationslope}The period difference as a function of period for $l = 1, m = -1$ modes in a 1.6 M$_{\odot}$ model with no convective overshoot.  As the rotation rate increases, thradiallye upward slope in the period spacing becomes more pronounced, and the change in slope occurs at lower periods.  }
\end{figure}

We find that rotation does not affect the overall level of the period spacing, but does affect the slope. For $m = 0$ modes, rotation causes the period difference to decrease with increasing period, similar to the results presented by \cite{vanreeth16,ouazzani2016}.  This makes the $m = 0$ modes appear more like $m = 1$ modes.  A similar trend is seen for the $m = 1$ modes, as rotation makes the negative slope steeper.  The $m = -1$ modes may provide a useful diagnostic of rotation in these stars, as shown in Figure \ref{fig:rotationslope}.  These modes show a clear upward trend at high periods in the nonrotating models.  As the rotation rate increases, the point at which the period difference starts to increase moves to lower periods, and the positive slope becomes steeper.  

\section{Conclusion}
\label{sec:conclusions}

We have calculated a grid of stellar models from 1.2 to 2.2 M$_{\odot}$ using MESA and GYRE.  Using these models, we have calculated the pulsation constant (Q) for the predicted $p$ modes.  We find that the previous fitting functions for Q published by \citet{Fitch81} work well above log$T \approx 3.83$.  Below this temperature, we find a sharp increase in the Q value for radial modes, and a large scatter toward smaller Q for non-radial modes.  This temperature range was outside of the grid of models used by \citet{Fitch81}, and so the trend is not accounted for by these fitting relations.  We find that the slope of the logQ-log$T_{eff}$ plot is strongly correlated with the amount of convective overshoot, and we provide fitting relations that can be used to determine the convective overshoot of a population of stars in a statistical sense.  We apply this method to a small sample of HADS known to pulsate radially and provide estimates for the convective overshoot.  This estimate has very large uncertainties, in part because of the assumptions used to estimate Q for these stars.  More accurate measurements as part of a dedicated study should be able to reduce these uncertainties and place tighter constraints on the convective overshoot.

We also find that the logQ-log$T_{eff}$ plot is relatively flat above log$T_{eff} \approx 3.83$, and that the level of this region is sensitive to the rotation rate.  As for convective overshoot, we provide a fitting function that can in principle be used to determine the rotation rate.  Unfortunately, our test sample had too few stars in this region to give a reliable determination of the average rotation rate of the sample.  

A comparison of our model Q values to the predictions for non-radial modes taken from \citet{Fitch81} do not agree as well as for the radial modes.  We find a bimodal distribution of Q values for the $\ell = 1$ modes, with the lower Q values corresponding to main sequence models, and the larger Q values corresponding to models evolving through the blue hook.  The predicted Q values lie in between these two sequences, but does not agree with either case.  

We calculate frequency and period spacing patterns for these stars, confirming previous research on the expected patterns.  Plotting echelle diagrams for the theoretical $p$ modes, we find that the large separation increases with both rotation rate and convective overshoot, but it is difficult to disentangle the effects.  Although it seems that the effects of rotation and overshoot can be seen in the echelle diagram of $\delta$ Scuti stars, the effects are degenerate and it is not possible to uniquely determine either parameter individually.  

We also plotted period spacings for the theoretical $g$ modes.  We confirm previous trends detected by \cite{ouazzani2016} and \cite{vanreeth15,vanreeth16}.  We also find that the $m = -1$ modes may provide a useful diagnostic of rotation rate, as the upward slope of the period spacings becomes steeper as the rotation rate increases.  We also find that the change in slope occurs at lower period as the rotation rate increases.
\acknowledgments

The authors would like to thank the anonymous referee for many helpful comments, which greatly improved this paper. This research was supported in part by the National Science Foundation under Grant No. NSF PHY-1125915.  This work was supported in part by NASA ATP grant 12-ATP12-0130.  The authors are grateful to the Kavli Institute at UCSB and the organizers of the Massive Star program for providing the opportunity to visit. This research has made use of NASA's Astrophysics Data System, the International Variable Star Index (VSX) database, operated at AAVSO, Cambridge, Massachusetts, USA, and the SIMBAD database, operated at CDS, Strasbourg, France.  Computational facilities are provided by ACENET, the regional high performance computing consortium for universities in Atlantic Canada. ACENET is funded by the Canada Foundation for Innovation (CFI), the Atlantic Canada Opportunities Agency (ACOA), and the provinces of Newfoundland and Labrador, Nova Scotia, and New Brunswick.

\bibliographystyle{yahapj}
\bibliography{main}

\end{document}